\documentclass[10pt,conference,a4paper]{elektron}
\usepackage[T1]{fontenc}
\usepackage[utf8]{inputenc}
\usepackage[compress]{cite}
\usepackage{graphicx}
\usepackage{xcolor}
\usepackage{comment}
\usepackage{balance}
\usepackage{multirow}
\usepackage{booktabs}
\usepackage{float}
\usepackage{hyperref}
\usepackage[caption=false]{subfig}
\usepackage{placeins}

\newcommand\blfootnote[1]{%
  \begingroup
  \renewcommand\thefootnote{}\footnote{#1}%
  \addtocounter{footnote}{-1}%
  \endgroup
}

\begin{document}

\title{A Comparative Study between HLS and HDL on SoC for Image Processing Applications \\ \vspace{0.5cm}
\large Un Estudio Comparativo entre HLS y HDL en SoC para Aplicaciones de Procesamiento de Imágenes}

\author{\IEEEauthorblockN{Roberto Millón\IEEEauthorrefmark{1}$^1$, Emmanuel Frati\IEEEauthorrefmark{1}$^2$ and Enzo Rucci\IEEEauthorrefmark{2}$^3$\vspace{0.2cm}}\IEEEauthorblockA{\IEEEauthorrefmark{1}\emph{Departamento de Ciencias Básicas y Tecnológicas, UNdeC}\\\emph{ Chilecito (5360), La Rioja, Argentina}}\IEEEauthorblockA{$^1$\texttt{\small{rmillon@undec.edu.ar}}\\$^2$\texttt{\small{fefrati@undec.edu.ar}}}\IEEEauthorblockA{\IEEEauthorrefmark{2}\emph{III-LIDI, Facultad de Informática, UNLP - CIC.}\\\emph{50 y 120 s/n, La Plata (1900), Argentina}\\$^3$\texttt{\small{erucci@lidi.info.unlp.edu.ar}}}}



\maketitle

\begin{abstract}

The increasing complexity in today's systems and the limited market times demand new development tools for FPGA. Currently, in addition to traditional hardware description languages (HDLs), there are high-level synthesis  (HLS) tools that increase the abstraction level  in system development.  Despite the greater simplicity of design and testing, HLS has some drawbacks in describing hardware.  This paper presents a comparative study between HLS and HDL for FPGA, using a Sobel filter as a case study in the image processing field. The results show that the HDL implementation is slightly better than the HLS version considering resource usage and response time. However, the programming effort required in the HDL solution is significantly larger than in the HLS counterpart.  
\\
\\
$~~$\emph{Resumen---} 
La creciente complejidad de los sistemas actuales y los tiempos limitados del mercado exigen nuevas herramientas de desarrollo para las FPGAs.  Hoy en día, además de los tradicionales lenguajes de descripción de hardware (HDL), existen herramientas de síntesis de alto nivel (HLS) que aumentan el nivel de abstracción en el desarrollo de sistemas.  A pesar de la mayor simplicidad de diseño y pruebas, HLS tiene algunos inconvenientes para describir hardware.  Este documento presenta un estudio comparativo entre HLS y HDL para FPGA, utilizando un filtro Sobel como caso de estudio en el ámbito del procesamiento de imágenes. Los resultados muestran que la implementación HDL es levemente mejor que la versión HLS considerando uso de recursos y tiempo de respuesta. Sin embargo, el esfuerzo de programación  en la implementación de HDL es significativamente mayor.
\\
\\
Palabras clave: FPGA; SoC; Sobel; HDL; HLS.
\blfootnote{
The final authenticated version is available online at \url{https://doi.org/10.37537/rev.elektron.4.2.117.2020}}
\end{abstract}


\IEEEpeerreviewmaketitle

\section{Introduction}
\label{sec:intro}



FPGAs are in an intermediate position between ASICs and CPUs, given their ability to reconfigure their architecture according to the application and their good energy efficiency~\cite{putnam_reconfigurable_2016}. 
In the last decade, multiple efforts have been made to reduce the energy consumption of large computing systems~\cite{Czarnul2019} and FPGAs are consolidating as a viable alternative to achieve this goal. That is why several companies and organizations have incorporated this kind of hardware devices to their systems, like Microsoft\cite{noauthor_microsoft_2018}, Baidu~\cite{noauthor_baidu_2017}, CERN~\cite{noauthor_cern_2017} or Amazon~\cite{noauthor_amazon_2017}.


However, FPGAs have not been massively adopted as it was originally expected by their vendors~\cite{tessier_reconfigurable_2015}. At the beginning and for more than one decade,  FPGA applications were exclusively developed using hardware description languages (HDLs). Unfortunately, HDLs have many drawbacks: they are verbose and error-prone, require in-depth knowledge of digital electronics, and demand long development times~\cite{zwagerman_high_2015}. As a result, the FPGA community explored alternative tools to increase the abstraction level as a way to reduce programming costs and accelerate time to market~\cite{windh_high-level_2015}.


Since the early 2000s, several FPGA vendors started to offer high-level synthesis (HLS) tools for systems development. In the HLS approach, programmers code FPGA applications using high-level languages (HLLs), like C, C++, or SystemC. Then, the tool is responsible for generating the corresponding HDL code~\cite{martin_high-level_2009}. Thus, engineers work at a higher abstraction level and produce reusable hardware designs without requiring hardware expertise. This alternative approach allowed the industry to shorten time to market since productivity gets increased while development cost gets reduced\cite{nane_survey_2016}.


Even though HLS tools present several advantages to develop hardware descriptions, they also have a weak spot. As HLLs were designed for software applications, they present some shortcomings  when describing hardware 
that can negatively impact the resources usage and response time of the final hardware designs~\cite{ren_brief_2014}.



In this context, it is important to know the advantages and disadvantages of different languages and approaches to synthesize optimal hardware descriptions.
This paper presents a comparative study between two Sobel filter solutions for a System-on-Chip (SoC) platform using both HDL and HLS approaches.   Overall, the main contributions of the paper are the following:
\begin{itemize}
    \item The creation of a public git repository containing optimized SoC solutions of Sobel filter for edge detection using both HDL and HLS approaches~\footnote{\url{https://github.com/robertoamt/HDL-HLS-Sobel-filters-}}. As Sobel is a convolutional operator, these implementations can be easily adapted to perform other image processing filters.
    \item A thorough comparison between both solutions in terms of resource usage, execution time, and programming effort. In this way, we can identify the strengths and weaknesses of each programming approach in the image processing field. 
\end{itemize}

The rest of the paper is organized as follows. Section \ref{sec:back} presents the background and the state of the art of this work. The optimized implementations are described
in Section~\ref{sec:imple}. In Section~\ref{sec:results}, experimental results are analyzed and finally, in Section~\ref{sec:conclusions}, conclusions and some ideas for
future research are summarized.

\section{Background and State of the Art}
\label{sec:back}

 \subsection{FPGA Programming Languages}
 
 
 Verilog and VHDL are the two leading HDLs to describe, simulate, and synthesize hardware systems. Both were developed in the 80's and have been updated several times since then~\cite{harris_digital_2013}.
 Using HDL to describe hardware requires digital design expertise, which limits the use of FPGA to hardware engineers. The results are low-level, complex designs, and slow development and debugging processes.
 
 
 The HDL drawbacks lead to the development of new tools to describe hardware in the early 2000s, such as Vivado HLS (Xilinx), Catapult C (Mentor Graphics), and Intel OpenCL SDK (Intel). These HLL-based tools raise the abstraction level and increase FPGA opportunities to engineers that specialize in embedded software programming~\cite{escobar_suitability_2016}. 
 
 Unfortunately, C-based languages have some shortcomings when describing hardware. First, either the designer or the HLS tool must specify the concurrency model because of HLLs lack a definition of hardware timing. Second, HLLs lack of the definition of exact bit width
for a signal, since they only provide limited data types such as bool, int, and/or long. Third, HLLs do not have abstractions of hardware interfaces and, contrary to the distributed memory model of FPGAs, they assume a flat memory model that can be accessed through pointers. Therefore, HLS tools must provide extensions to the HLL through libraries and directive sets to overcome those deficiencies. In other cases, HLS tools impose restrictions on HLLs, such as not supporting dynamic memory allocations~\cite{ren_brief_2014}.
 
  \subsection{Sobel Filter}


The Sobel algorithm is a gradient-based edge detection method to extract the edges of a grayscale image using the first derivative. By computing horizontal $h$ and vertical $v$ direction derivatives of a pixel against the surrounding pixels, the algorithm segments the image into areas or objects. This process reduces the amount of data while preserving the image's structural properties.
The $G_h$ and $G_v$ derivatives represent the components of the gradient vector~\cite{vallina_zynq_2012,zheng_design_2017}, given by Equation~\ref{eq:gradiente}.

\begin{equation}
\nabla f = 
\begin{pmatrix}
{    
G_{h},
G_{v}
}
\end{pmatrix}
\label{eq:gradiente}
\end{equation}


The gradient magnitude expresses the rate of change of intensity in neighboring pixels and defines the edge strength. A sudden change in contiguous pixels increases the gradient magnitude resulting in the border of an object, given by Equation~\ref{eq:rms}:

\begin{equation}
|\nabla f| = \sqrt{{G_{h}}^2 + {G_{v}}^2}  
\label{eq:rms}
\end{equation}

Equation~\ref{eq:rms} can be approximated as Equation~\ref{eq:abs}. This last formula delivers a faster computation but still preserving relative changes in intensity.

\begin{equation}
|\nabla f| \approx |G_{h}| + |G_{v}| 
\label{eq:abs}
\end{equation}

\begin{figure}[ht]
\centering
\includegraphics[width=0.7\columnwidth]{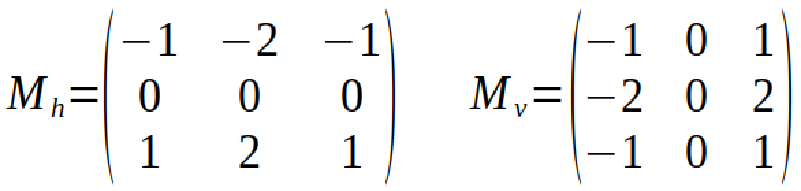}
\caption{Sobel convolution masks}
\label{fig:mascaras}
\end{figure}

Fig.~\ref{fig:mascaras} shows the two $3\times3$ convolution masks (namely, $M_h$ and $M_v$) used by the Sobel filter to calculate the components of the gradient vector. 

\begin{figure}[h]
    \centering
    \includegraphics[width=0.8\columnwidth]{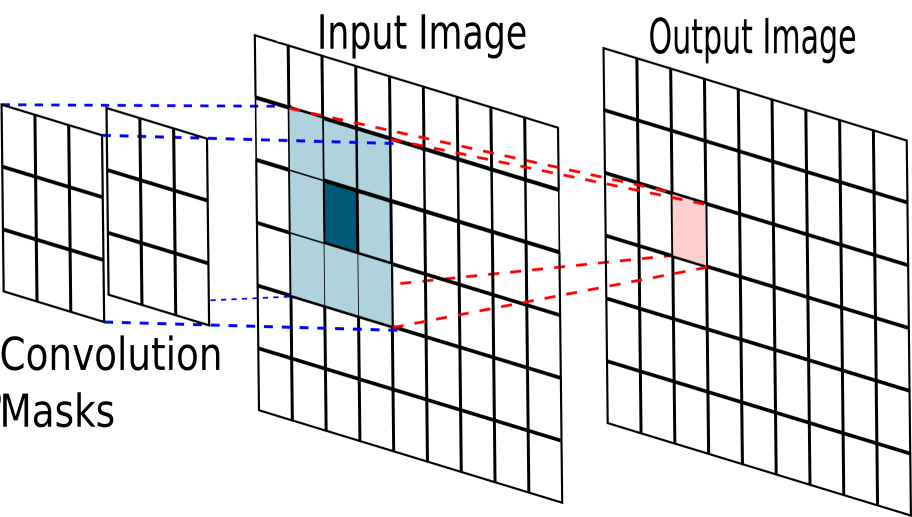}
    \caption{Sobel convolution process}
    \label{fig:conv2}
\end{figure}

From mathematical point of view,  the image must be multiplicated by the Sobel masks to get the components of the gradient vector. The image is scanned from left to right and top to bottom, applying convolution to each individual pixel using the $M_h$ and $M_v$ masks~\cite{nausheen_fpga_2018}. Fig.~\ref{fig:conv2} shows the convolution process of the Sobel filter.

\subsection{State of the Art}
 
In the literature, numerous works propose
HLS or HDL solutions to different problems. However,
just a few implement the same algorithm with both HLS
and HDL descriptions.  
Some studies conclude that both implementations have similar performance results with a larger resource consumption on the HLS approach~\cite{hiraiwa_fpga_2013,zwagerman_high_2015,hill2015,wang2015,stanciu_comparison_2017,stamoulias_hardware_2017}. 
Conversely, other studies observed better performances in one of the two implementations, either HLS~\cite{Cornu2011,Bachrach2012,arcas2014,gurel2016,pelcat_design_2016} or HDL~\cite{tetrault_two_2018,akkad2018}. 
From the programming effort perspective, comparative works between HLS and HDL designs for FPGA show similar trends (except for~\cite{aledo_vhdl_2019}). In general, HLS descriptions require less development time due to their higher abstraction level and the programmer's \textit{familiarity} with those languages. However, there is no single conclusion about performance and resource usage between the two approaches. In these aspects, the results are affected by the problem characteristics, the tools used, and the design features.
 
This research presents and compares two optimized Sobel filter solutions for a SoC platform designed with both HLS and HDL approaches. 
Compared to previous works, the novelty of this study lies in the hardware and software technologies used, the chance to reuse/modify the solutions implemented, and the careful comparative analysis carried out.
In this way,  we can contribute to the identification of the strengths and weaknesses of each programming approach in the image processing field.

\section{Implementation}
\label{sec:imple}

\begin{figure}[ht]
\centering
\includegraphics[width=0.8\columnwidth]{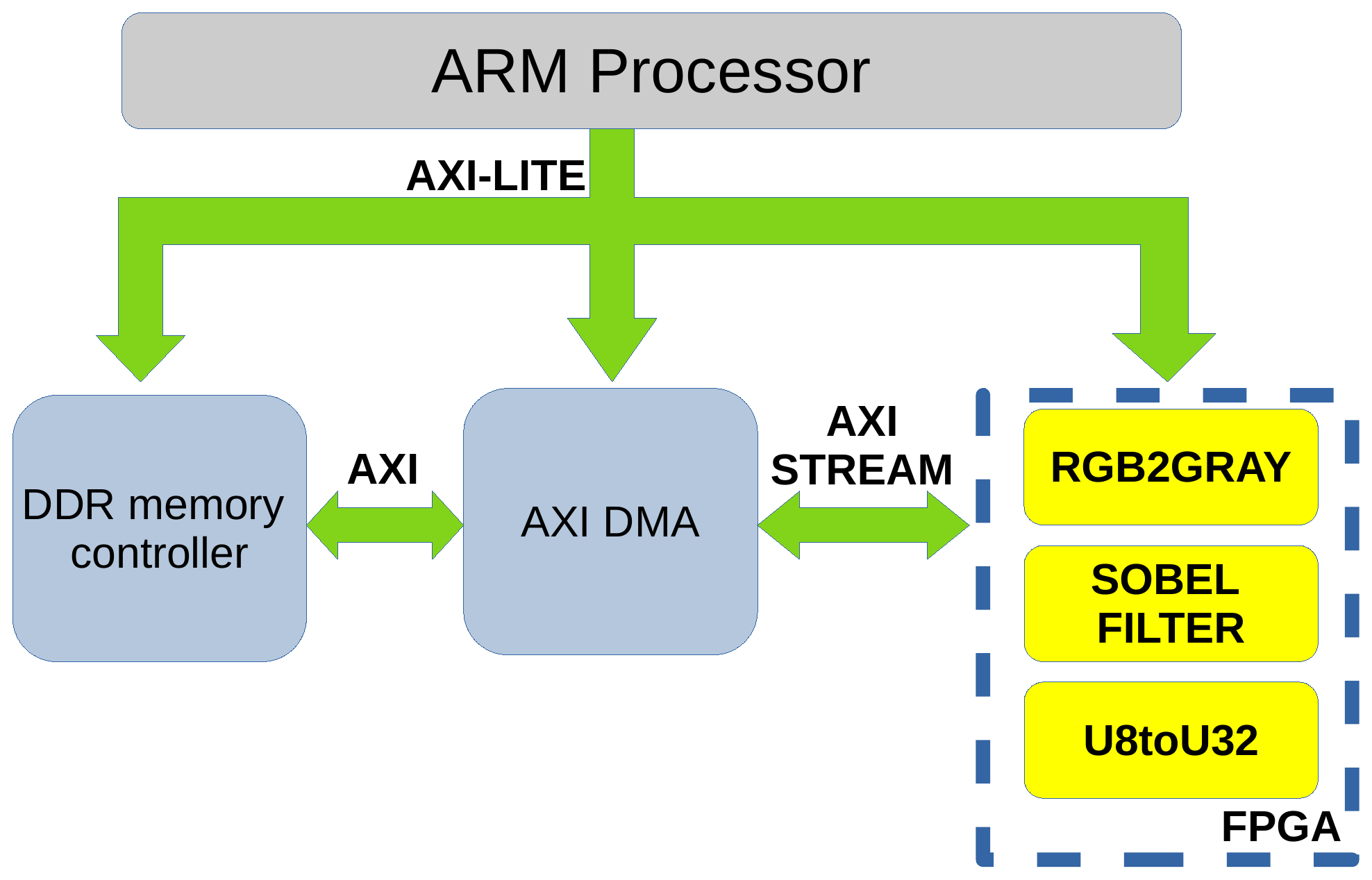}
\caption{Complete Sobel system}
\label{fig:system}
\end{figure}

To compare both HDL and HLS approaches, we have implemented a complete Sobel system to detect edges on RGB images (BMP format). Firstly, a DMA block accesses the microSD memory for image reading and sends the pixel stream to the processing core. After Sobel operation, the processing block returns the pixel stream to the DMA block for further storage in the microSD memory. The ARM processor is responsible for configuring and managing the system. Finally, all modules were integrated using the Vivado 2019.1 tool. Fig.~\ref{fig:system} shows a block diagram of the whole system.

The processing core consists of 3 blocks that were synthesized and validated using both HDL and HLS approaches.
The first block (RGB2GRAY) converts RGB images to grayscale using the method based on the arithmetic mean of the three components; the second block (SOBEL FILTER) detects the edges in the image; and the third block (U8toU32) combines four 8-bit integer variables to form a 32-bit word.


\subsection{HLS Version}
\label{subsec:func}


To implement the  Sobel filter on HLS, we use two memory structures called \emph{line buffer} and \emph{sliding window} ~\cite{millonCASE2020}. Each line buffer correspond to an entire row of the image and we used them to keep the Sobel working set. On its behalf, the sliding window contains the image pixels that will be convoluted. 

\begin{figure}[ht]
\centering
\includegraphics[width=1\columnwidth]{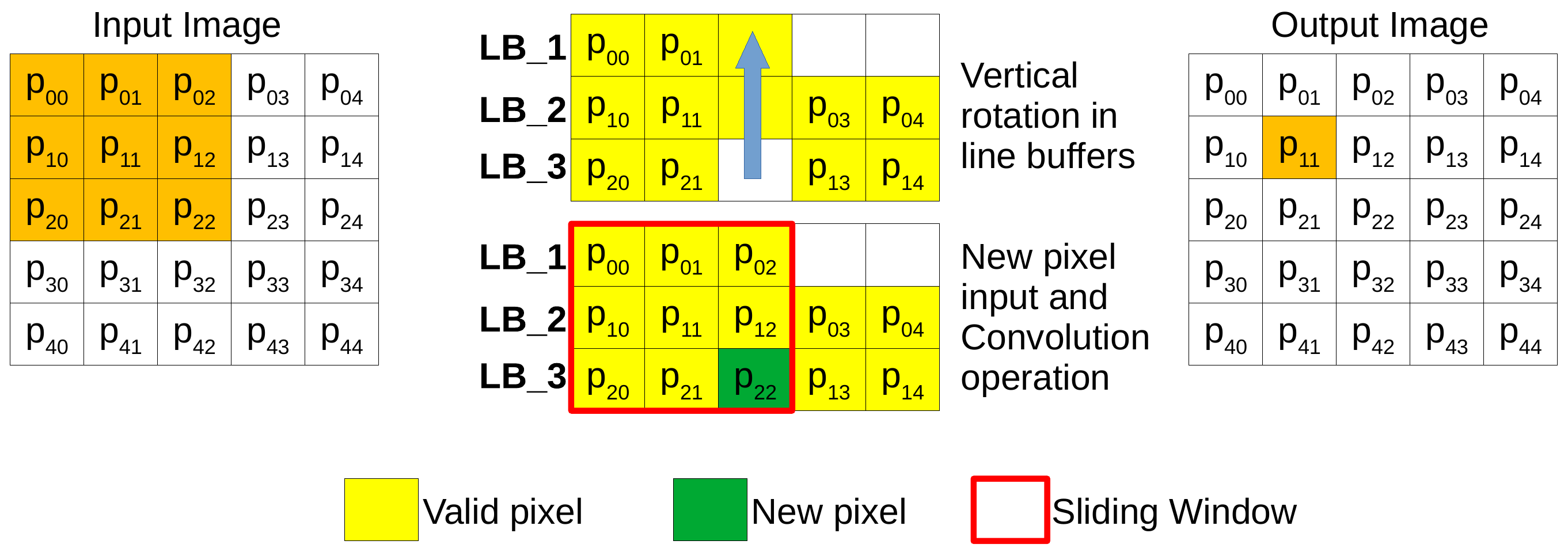}
\caption{Sobel filter operation in the HLS implementation}
\label{fig:convSobel}
\end{figure}

The filter requires three line buffers that are arranged one above the other. When a new pixel is received, the line buffers perform a vertical rotation before storing the new pixel. When the two upper line buffers are full and three pixels are in the lowest line buffer, the convolution process is performed using the sliding window pixels and the Sobel masks. Fig~\ref{fig:convSobel} illustrates the described process. 

Several compiler directives were used to optimize synthesis and reduce programming costs. To communicate the image and its parameters, the system uses AXI4 interfaces that were synthesized using the \texttt{HLS INTERFACE} pragma. The design also incorporates two compiler directives to achieve better performance: \texttt{HLS ARRAY PARTITION} pragma allows parallel memory accesses to line buffers, sliding window and masks; and the \texttt{HLS PIPELINE} pragma to increase throughput by enabling parallel execution of convolution tasks.

\subsection{HDL Version}

\begin{figure}[ht]
\centering
\includegraphics[width=.9\columnwidth]{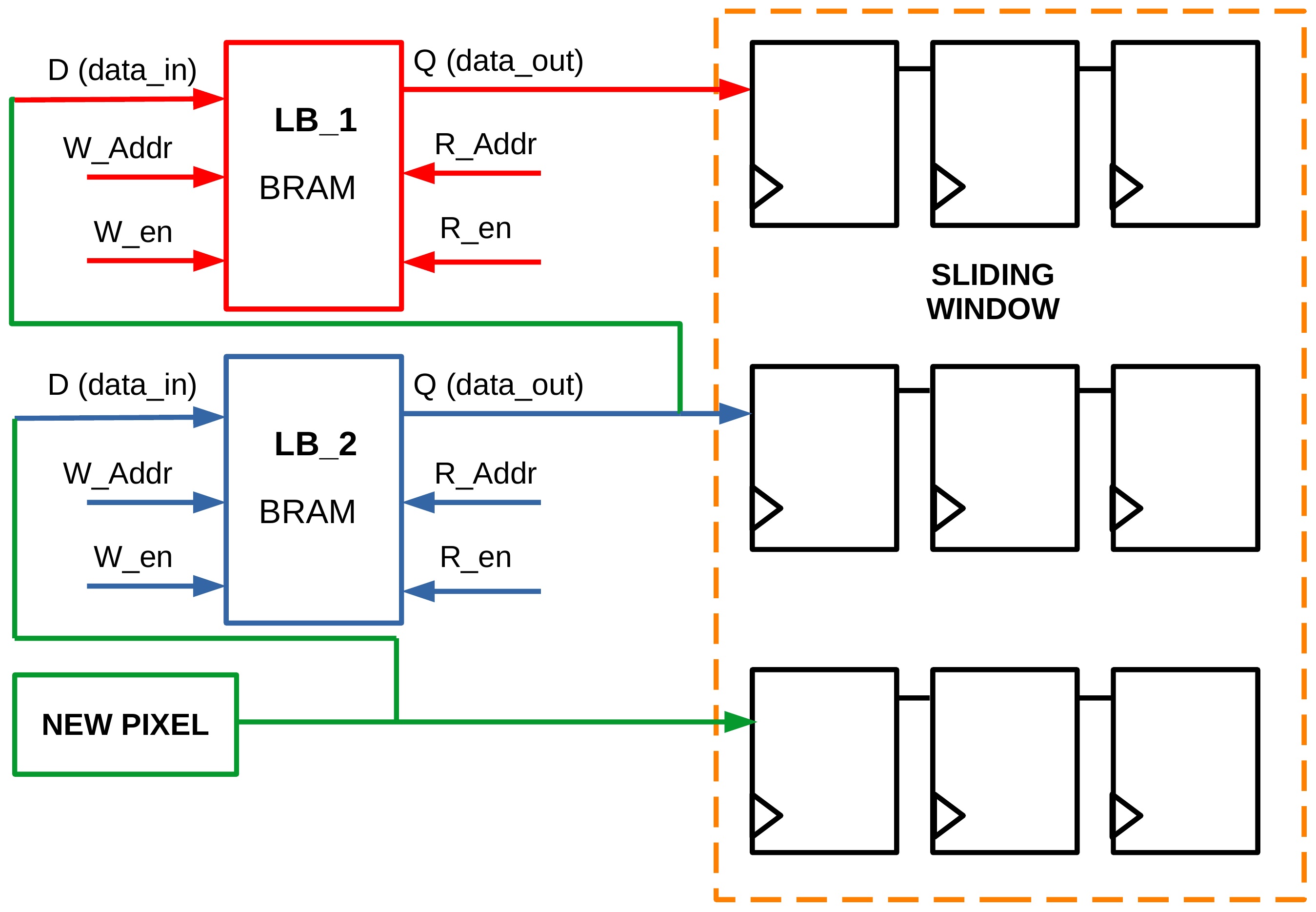}
\caption{Synthesis of Sobel filter in HDL implementation}
\label{fig:vhdlsys}
\end{figure}

As the HLS version, the HDL implementation also uses line buffers and sliding window data structures. However, it just requires two line buffers and not three of them. Previous HDL solutions~\cite{nosrat_hardware_2012,mehra_area_2012,chaple_design_2014} use shift registers to describe the line buffers, which results in a larger use of FPGA registers. To decrease the use of registers, our design uses two blocks of RAM RAMB18 (one physical RAMB36) as line buffers (namely  \textit{line\_buffer\_1} and \textit{line\_buffer\_2}). We synthesized the sliding window as three shift registers composed of three flip-flops each (9 registers in all). Fig~\ref{fig:vhdlsys} represents the HDL implementation of Sobel filter. 


\begin{figure}[ht]
\centering
\includegraphics[width=1\columnwidth]{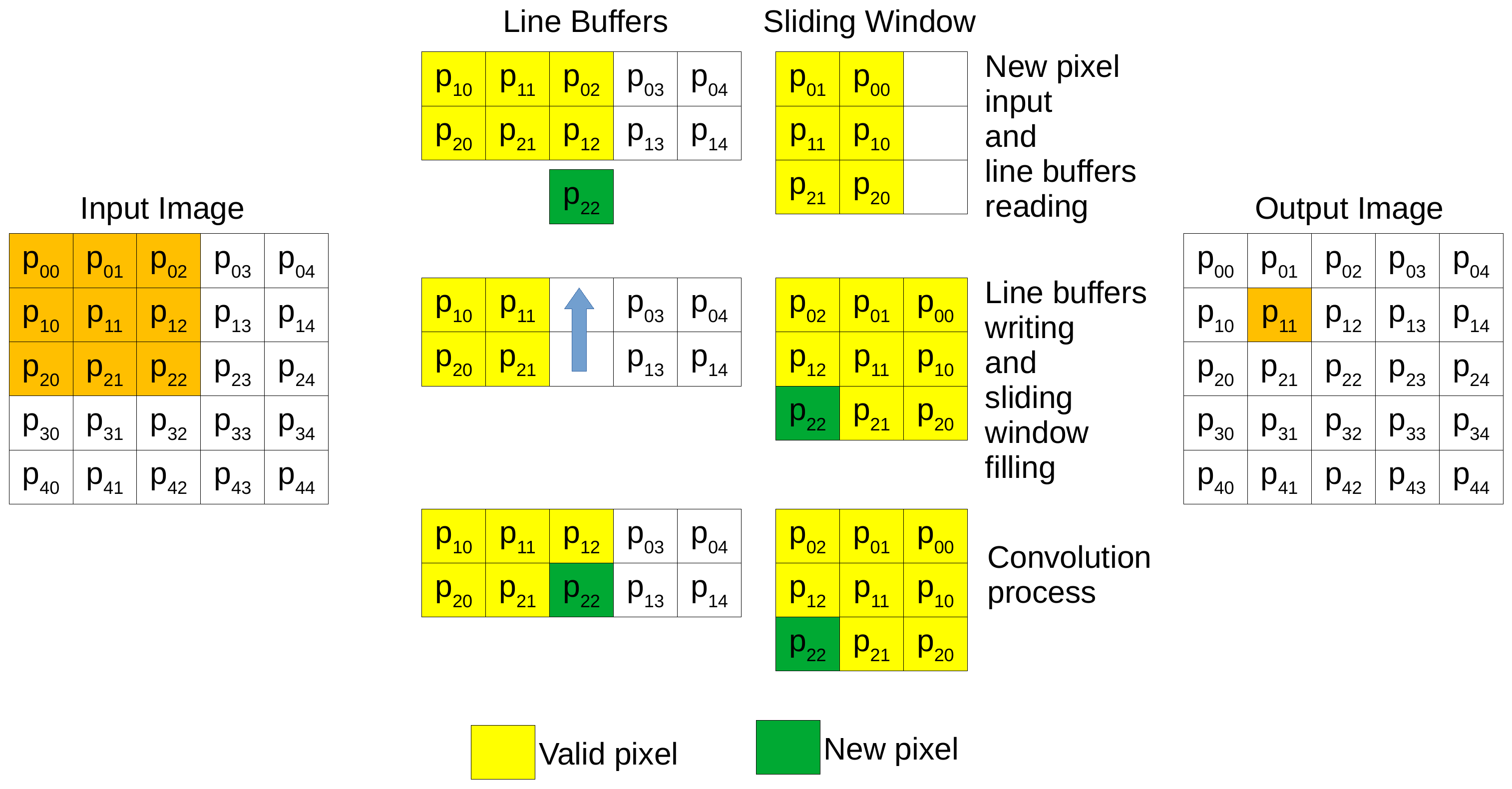}
\caption{Sobel filter process in HDL implementation}
\label{fig:hdl_tasks}
\end{figure}

Sobel filter performs four tasks, each requiring one clock cycle. The first task is comprised of two actions: the reception of a new pixel and the reading of the line buffers. 
The second task is composed of two actions too: the filling of the sliding windows and the writing in the line buffers. In the next clock cycle, the filter performs the convolution. Finally, the filter sends the processed data in the fourth cycle. This process is shown in Fig.~\ref{fig:hdl_tasks}.

\begin{figure}[ht]
\centering
\includegraphics[width=.9\columnwidth]{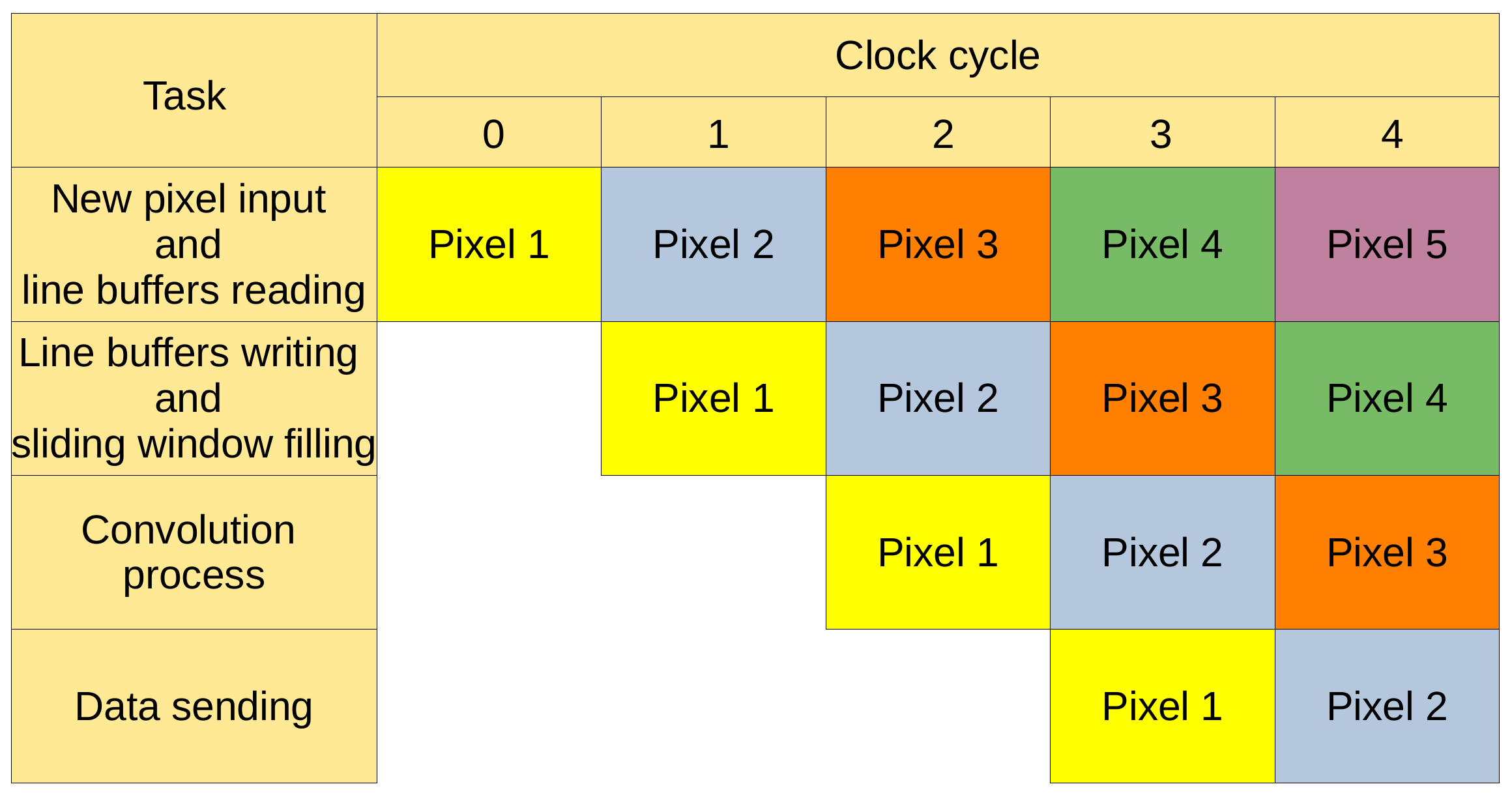}
\caption{Pipelined scheme in HDL implementation}
\label{fig:pipeline Tasks}
\end{figure}

Unlike the HLS design, the HDL system must describe the AXI4 connection interfaces at a lower level (AXI-STREAM and AXI-LITE). As it was mentioned before, the complete process was divided into four stages, incorporating registers between them. Although these additional registers increase the latency, they allow this design to  overlap the different tasks in the same clock cycle. In this way, a pipelining scheme is manually achieved (no compiler directives), which increases system throughput and shortens response time. The pipelined tasks are shown in Fig.~\ref{fig:pipeline Tasks}.

\section{Experimental Results}
\label{sec:results}
The SoC used for testing is a ZYBO platform (SoC ZYNQ-7000), which consists of an ARM Cortex-A9 dual-core processor and an XC7Z010-1-CLG400C FPGA. To system design, we have used the graphical environment within the Vivado Design Suite software. 
We developed a test application using the XSDK tool and also selected four images from public repositories: \textit{Mandrill} of 512$\times$512\footnote{\url{http://sipi.usc.edu/database/database.php}}, \textit{Kodim23} of 768$\times$512\footnote{\url{http://www.cs.albany.edu/~xypan/research/img/Kodak/kodim23.png}}, \textit{Owl} of 1920$\times$566\footnote{\url{https://pixabay.com/es/photos/lechuza-granero-ave-animales-1710659/}}, and \textit{Lightbulbs} of 1920$\times$1080\footnote{\url{https://pixabay.com/es/illustrations/bombillas-de-luz-para-vidrio-5488573/}}. Each particular test was run ten times; performance was calculated by the average of ten
executions to avoid variability. All time measures were performed with the \texttt{xtime\_l.h} library.

\subsection{Edge Detection}

\begin{figure}[!htbp]
    \centering
    \subfloat[]{{\includegraphics[width=.95\columnwidth]{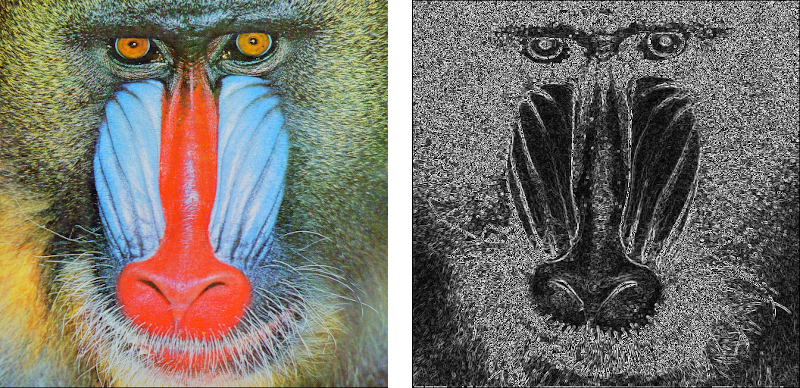} }}%
    \\
    \subfloat[]{{\includegraphics[width=.95\columnwidth]{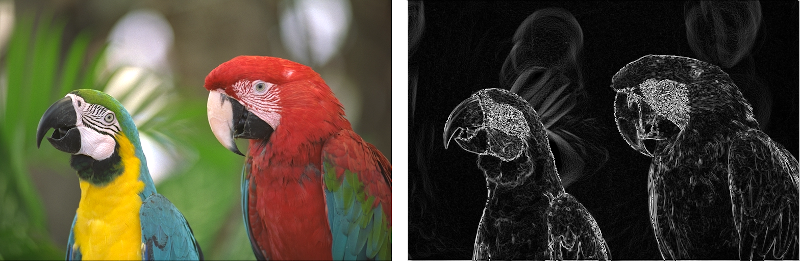} }}%
    \\
    \subfloat[]{{\includegraphics[width=.95\columnwidth]{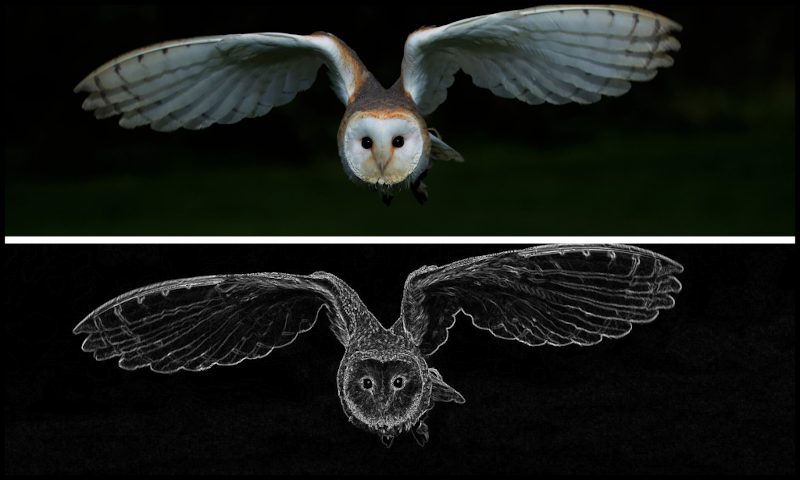} }}%
    \\
    \subfloat[]{{\includegraphics[width=.95\columnwidth]{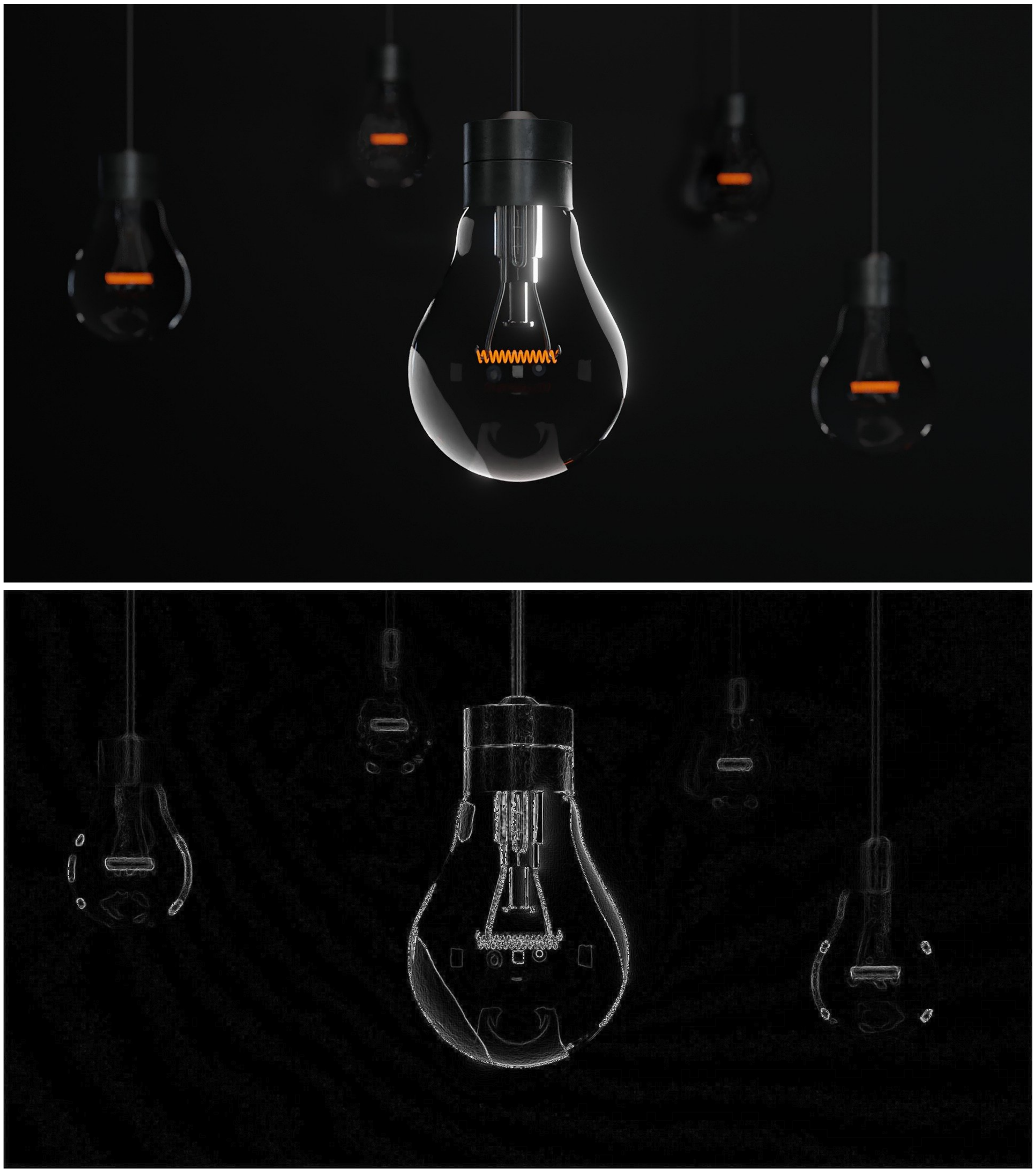} }}%
    \caption{Test images. (a) \textit{Mandrill}. (b) \textit{Kodim23}. (c) \textit{Owl}. (d) \textit{Lightbulbs}.}   
    \label{fig:fig_results}
\end{figure}

Fig.~\ref{fig:fig_results} shows the testing images with their corresponding resulting image after applying the Sobel filter. The approximated formula from Equation~\ref{eq:abs} was used to compute the gradient magnitude. Last, we have compared the resulting images of both approaches (HDL \& HLS) using an automated procedure based on the Hamming distance. No difference between images were found in all cases.

\subsection{Resource Usage and Performance} 

Table~\ref{tab:resource_table} presents the resource usage for the two Sobel implementations. The values in the columns \texttt{S.LUTs}, \texttt{S.Registers}, \texttt{F7 Muxes}, \texttt{BRAM}, and \texttt{DSPs} refer to the percentages of lookup tables, registers, multiplexers, RAM blocks, and DSP blocks used, respectively. 
In particular, the HLS version requires 5.5$\times$ more LUTs and 4.6$\times$ more registers than the HDL counterpart. Still, it does not represent a design restriction in either case. 

\begin{table}[!htbp]
\centering
\caption{Resource usage}
\resizebox{\columnwidth}{!}{%
\begin{tabular}{|c|c|c|c|c|c|}
\toprule
\multirow{2}{*}{\textbf{Version}} &
\multicolumn{5}{c|}{{\textbf{Resource usage (\%)}}} \\
\cmidrule(l){2-6}  & S.LUTs & S.Reg & F7 Muxes & BRAM & DSPs \\
\midrule
HDL & 0.8\% & 0.5\% & 0\% & 1.6\% & 0\% \\
\midrule
HLS & 4.4\% & 2.3\% & $<$ 0.1\% & 1.6\% & 0\% \\
\bottomrule
\end{tabular}%
}
\label{tab:resource_table}
\end{table}

Table ~\ref{tab:timing_table} presents the runtime of each Sobel version. For each testing image, we present two times: (1) the operation time of the processing blocks (RGB2GRAY, SOBEL FILTER, and U8toU32) labeled as \textit{Sobel}, and (2) the I/O time of the system labeled as \textit{I/O}. 
Also, the last column shows the speedup ratio between the HLS runtime and the HDL runtime~\footnote{This value is only computed for the Sobel column of each image. The I/O time is approximately the same for both versions since they share the I/O operation}.
As it can be seen in the results, the HDL filter is faster than the HLS version in all cases. The largest performance difference is achieved when processing the smallest image (\textit{Mandrill}), reaching a speedup of 6.6$\times$. The difference decreases as the size of the image increases, but the HDL version still shows superior to the HLS counterpart obtaining a speedup of 1.4$\times$ for the largest image (\textit{Lightbulbs}). 




\begin{table}[!htbp]
\centering
\caption{Performance}
\resizebox{\columnwidth}{!}{
\begin{tabular}{|c|c|c|c|c|}
\toprule
 \multirow{2}{*}{\textbf{Image}} & \multirow{2}{*}{\textbf{Task}} & \multicolumn{3}{c|}{{\textbf{Runtime (ms)}}} \\ \cmidrule(r){3-5}
& & HDL & HLS & Speedup \\ \midrule
\textbf{Mandrill} & \textbf{Sobel} & 5.9 & 39.3 & 6.6$\times$ \\ \cmidrule(r){2-5}
(512$\times$512) & \textbf{I/O} & 816 & 837 & - \\ \midrule
\textbf{Kodim23} & \textbf{Sobel} & 8.8 & 39.3 & 4.4$\times$ \\ \cmidrule(r){2-5}
(768$\times$512) & \textbf{I/O} & 1078 & 1026 & - \\ \midrule
\textbf{Owl} & \textbf{Sobel} & 24.4 & 43.4 & 1.7$\times$ \\ \cmidrule(r){2-5}
(1920$\times$566) & \textbf{I/O} & 2280.4 & 2329.5 & - \\ \midrule
\textbf{Lightbulbs} & \textbf{Sobel} & 58 & 82.9 & 1.4$\times$ \\ \cmidrule(r){2-5}
(1920$\times$1080) & \textbf{I/O} & 3985.9 & 3954.5 & - \\ \bottomrule
\end{tabular}
}
\label{tab:timing_table}
\end{table}


The HDL implementation outperformed the HLS version for both resource usage and performance. Both issues are related to the automated process that Vivado HLS follows to translate the HLL code to the corresponding hardware description. The resulting RTL requires a larger number of finite state machines and signals to ensure its correct operation, increasing resource consumption and response time. 


\subsection{Programming Cost}

There are several proposals for measuring the programming cost of an application. Some of them propose counting the number of lines of code (SLOC) or the number of characters (including blank lines and comments). Despite their simplicity, the main drawback is that these parameters do not reflect the complexity of the algorithms~\cite{bhatt_analysis_2012}. Other alternatives measure the development time, even though it is dependent on the  programmer experience. 
In this work, we have decided to measure the programming cost through the SLOC indicator and the development time invested to reach a complete and functional implementation. These metrics are combined with a qualitative comparison of the required effort in each solution. As both parts are complementary, they allow the reader a comprehensive understanding of the programming cost.

Table~\ref{tab:progcost_table} shows the number of files, SLOC~\footnote{To measure the SLOC indicator, we have used the cloc tool ~\url{https://github.com/AlDanial/cloc}}, and development time (in hours) for each approach.
The project involved a single programmer having minimal knowledge about FPGA programming~\footnote{The programmer took two undergraduate semester courses in the field.}. The HLS version was developed first, taking him 121 hours and 90 SLOC to get a correct implementation of the entire system. Next, the programmer continued with the HDL implementation. It is important to remark that, implementing the HLS version first gave some advantage to the HDL implementation due to the knowledge gained in that process.  Despite that, the development time increased to 384 hours and 493 SLOC.
This represents an increase of 5.5$\times$ SLOC and 3.2$\times$ hours compared to the HLS version.

\begin{table}[!htbp]
\centering
\caption{Programming cost}
\resizebox{.8\columnwidth}{!}{%
\begin{tabular}{|c|c|c|c|}
\toprule
\multirow{3}{*}{\textbf{Version}} &
\multicolumn{3}{c|}{{\textbf{Programming cost}}} \\\cmidrule(l){2-4}  
& \multirow{2}{*}{\# Files} & \multirow{2}{*}{SLOC} & Development \\
& & & time (hours) \\ \midrule
HDL & 4 & 493 & 384\\ \midrule
HLS & 2 & 90 & 121\\ \bottomrule
\end{tabular}%
}
\label{tab:progcost_table}
\end{table}

The HDL implementation required more time and SLOC than its HLS counterpart. This is due to the HLS approach allowed the programmer to focus on the system functionality without requiring him to define hardware resources and/or other low-level mechanisms. For example,  communication ports and pipelining scheme were implemented using compiler directives in the HLS approach. In the opposite direction, the programmer had to manually describe them at RTL level in the HDL approach.  
\FloatBarrier
\section{Conclusions and Future Work}
\label{sec:conclusions}

In this work, we present a comparative study beween HLS and HDL approaches for FPGA programming. Taking the Sobel filter as a study case, we implemented optimized SoC solutions for detecting edges in images and made them available through a public git repository. Next, we performed a thorough comparison  between HDL and HLS in terms of resource usage, execution time, and programming effort. As Sobel is a convolutional operator like many others, we can identify   strengths and weaknesses of each programming approach in the image processing field. 


The results show that the HDL implementation is slightly better than the HLS version considering resource usage and response time. However, the programming effort required in the HDL solution is significantly larger than in the HLS counterpart. According to these results, the HDL approach would only be convenient when the resource usage and/or the response times are critical. Otherwise, the HLS approach can lead to important reductions in both programming cost and development time, at the cost of a small increase in resource usage and execution time.

Future work focuses on extending the experimental work carried out to other boards. This would allow us to enlarge the representativity of the analysis performed.

\section*{Acknowledgment}

This work was partially supported by the ``Software y aplicaciones en computación de altas prestaciones'' project, RR Nº 883/18 (UNdeC). 



%

\bibliographystyle{IEEEtran}
\balance
\bibliography{references}

\begin{thebibliography}{10}
\providecommand{\url}[1]{#1}
\csname url@samestyle\endcsname
\providecommand{\newblock}{\relax}
\providecommand{\bibinfo}[2]{#2}
\providecommand{\BIBentrySTDinterwordspacing}{\spaceskip=0pt\relax}
\providecommand{\BIBentryALTinterwordstretchfactor}{4}
\providecommand{\BIBentryALTinterwordspacing}{\spaceskip=\fontdimen2\font plus
\BIBentryALTinterwordstretchfactor\fontdimen3\font minus
  \fontdimen4\font\relax}
\providecommand{\BIBforeignlanguage}[2]{{%
\expandafter\ifx\csname l@#1\endcsname\relax
\typeout{** WARNING: IEEEtran.bst: No hyphenation pattern has been}%
\typeout{** loaded for the language `#1'. Using the pattern for}%
\typeout{** the default language instead.}%
\else
\language=\csname l@#1\endcsname
\fi
#2}}
\providecommand{\BIBdecl}{\relax}
\BIBdecl

\bibitem{putnam_reconfigurable_2016}
\BIBentryALTinterwordspacing
A.~Putnam, J.~Gray, M.~Haselman, S.~Hauck, S.~Heil, A.~Hormati, J.-Y. Kim,
  S.~Lanka, J.~Larus, E.~Peterson, S.~Pope, A.~M. Caulfield, A.~Smith,
  J.~Thong, P.~Y. Xiao, D.~Burger, E.~S. Chung, D.~Chiou, K.~Constantinides,
  J.~Demme, H.~Esmaeilzadeh, J.~Fowers, and G.~P. Gopal,
  ``\BIBforeignlanguage{en}{A reconfigurable fabric for accelerating
  large-scale datacenter services},''
  \emph{\BIBforeignlanguage{en}{Communications of the ACM}}, vol.~59, no.~11,
  pp. 114--122, Oct. 2016. [Online]. Available:
  \url{http://dl.acm.org/citation.cfm?doid=3013530.2996868}
\BIBentrySTDinterwordspacing

\bibitem{Czarnul2019}
\BIBentryALTinterwordspacing
P.~Czarnul, J.~Proficz, and A.~Krzywaniak, ``Energy-aware high-performance
  computing: Survey of state-of-the-art tools, techniques, and environments,''
  \emph{Scientific Programming}, vol. 2019, p. 8348791, Apr 2019. [Online].
  Available: \url{https://doi.org/10.1155/2019/8348791}
\BIBentrySTDinterwordspacing

\bibitem{noauthor_microsoft_2018}
\BIBentryALTinterwordspacing
P.~Hernandez, ``Microsoft {Uses} {Intel} {FPGAs} for {Smarter} {Bing}
  {Searches},'' 2018. [Online]. Available:
  \url{https://www.eweek.com/cloud/microsoft-uses-intel-fpgas-for-smarter-bing-searches}
\BIBentrySTDinterwordspacing

\bibitem{noauthor_baidu_2017}
X.~Inc., ``Baidu {Deploys} {Xilinx} {FPGAs} in {New} {Public} {Cloud}
  {Acceleration} {Services}.''

\bibitem{noauthor_cern_2017}
\BIBentryALTinterwordspacing
L.~Barney, ``\BIBforeignlanguage{en-US}{{CERN} openlab {Explores} {New}
  {CPU}/{FPGA} {Processing} {Solutions}},'' Apr. 2017. [Online]. Available:
  \url{https://www.hpcwire.com/2017/04/14/xeon-fpga-processor-tested-at-cern/}
\BIBentrySTDinterwordspacing

\bibitem{noauthor_amazon_2017}
\BIBentryALTinterwordspacing
K.~Freund, ``Amazon {And} {Xilinx} {Deliver} {New} {FPGA} {Solutions},'' 2017.
  [Online]. Available: \url{https://bit.ly/3ofmBsI}
\BIBentrySTDinterwordspacing

\bibitem{tessier_reconfigurable_2015}
R.~Tessier, K.~Pocek, and A.~DeHon, ``Reconfigurable {Computing}
  {Architectures},'' \emph{Proceedings of the IEEE}, vol. 103, no.~3, pp.
  332--354, Mar. 2015.

\bibitem{zwagerman_high_2015}
M.~D. Zwagerman, ``High level synthesis, a use case comparison with hardware
  description language,'' mastersthesis, School of Engineering, Grand Valley
  State University, 2015.

\bibitem{windh_high-level_2015}
\BIBentryALTinterwordspacing
S.~Windh, X.~Ma, R.~J. Halstead, P.~Budhkar, Z.~Luna, O.~Hussaini, and W.~A.
  Najjar, ``\BIBforeignlanguage{en}{High-{Level} {Language} {Tools} for
  {Reconfigurable} {Computing}},'' \emph{\BIBforeignlanguage{en}{Proceedings of
  the IEEE}}, vol. 103, no.~3, pp. 390--408, Mar. 2015. [Online]. Available:
  \url{http://ieeexplore.ieee.org/document/7086410/}
\BIBentrySTDinterwordspacing

\bibitem{martin_high-level_2009}
G.~Martin and G.~Smith, ``High-level synthesis: Past, present, and future,''
  \emph{IEEE Des. Test}, vol.~26, no.~4, pp. 18--25, 2009, conference Name:
  {IEEE} Design Test of Computers.

\bibitem{nane_survey_2016}
R.~Nane, V.-M. Sima, C.~Pilato, J.~Choi, B.~Fort, A.~Canis, Y.~T. Chen,
  H.~Hsiao, S.~Brown, F.~Ferrandi, J.~Anderson, and K.~Bertels, ``A survey and
  evaluation of {FPGA} high-level synthesis tools,'' \emph{IEEE Transactions on
  Computer-Aided Design of Integrated Circuits and Systems}, vol.~35, no.~10,
  pp. 1591--1604, 2016.

\bibitem{ren_brief_2014}
\BIBentryALTinterwordspacing
H.~Ren, ``A brief introduction on contemporary {High}-{Level} {Synthesis},'' in
  \emph{2014 {IEEE} {International} {Conference} on {IC} {Design} \&
  {Technology}}.\hskip 1em plus 0.5em minus 0.4em\relax Austin, TX, USA: IEEE,
  May 2014, pp. 1--4. [Online]. Available:
  \url{http://ieeexplore.ieee.org/document/6838614/}
\BIBentrySTDinterwordspacing

\bibitem{harris_digital_2013}
D.~Harris and S.~Harris, \emph{\BIBforeignlanguage{en}{Digital {Design} and
  {Computer} {Architecture} (Second Edition)}}.\hskip 1em plus 0.5em minus
  0.4em\relax Morgan Kaufmann, 2013.

\bibitem{escobar_suitability_2016}
\BIBentryALTinterwordspacing
F.~A. Escobar, X.~Chang, and C.~Valderrama, ``Suitability {Analysis} of {FPGAs}
  for {Heterogeneous} {Platforms} in {HPC},'' \emph{IEEE Transactions on
  Parallel and Distributed Systems}, vol.~27, no.~2, pp. 600--612, Feb. 2016.
  [Online]. Available: \url{http://ieeexplore.ieee.org/document/7051218/}
\BIBentrySTDinterwordspacing

\bibitem{vallina_zynq_2012}
\BIBentryALTinterwordspacing
F.~M. Vallina, C.~Kohn, and P.~Joshi, ``Zynq all programmable {SoC} sobel
  filter implementation using the vivado {HLS} tool,'' 2012. [Online].
  Available: \url{https://bit.ly/3h6egD1}
\BIBentrySTDinterwordspacing

\bibitem{zheng_design_2017}
\BIBentryALTinterwordspacing
Y.~Zheng, ``The design of sobel edge extraction system on {FPGA},'' \emph{ITM
  Web Conf.}, vol.~11, 2017. [Online]. Available:
  \url{https://doi.org/10.1051/itmconf/20171108001}
\BIBentrySTDinterwordspacing

\bibitem{nausheen_fpga_2018}
N.~Nausheen, A.~Seal, P.~Khanna, and S.~Halder, ``\BIBforeignlanguage{en}{A
  {FPGA} based implementation of {Sobel} edge detection},''
  \emph{\BIBforeignlanguage{en}{Microprocess. Microsyst.}}, vol.~56, p.
  84–91, Feb. 2018.

\bibitem{hiraiwa_fpga_2013}
J.~Hiraiwa and H.~Amano, ``An {FPGA} {Implementation} of {Reconfigurable}
  {Real}-{Time} {Vision} {Architecture},'' in \emph{2013 27th {International}
  {Conference} on {Advanced} {Information} {Networking} and {Applications}
  {Workshops}}, Mar. 2013, pp. 150--155.

\bibitem{hill2015}
K.~{Hill}, S.~{Craciun}, A.~{George}, and H.~{Lam}, ``Comparative analysis of
  opencl vs. hdl with image-processing kernels on stratix-v fpga,'' in
  \emph{2015 IEEE 26th International Conference on Application-specific
  Systems, Architectures and Processors (ASAP)}, 2015, pp. 189--193.

\bibitem{wang2015}
G.~{Wang}, H.~{Lam}, A.~{George}, and G.~{Edwards}, ``Performance and
  productivity evaluation of hybrid-threading hls versus hdls,'' in \emph{2015
  IEEE High Performance Extreme Computing Conference (HPEC)}, 2015, pp. 1--7.

\bibitem{stanciu_comparison_2017}
\BIBentryALTinterwordspacing
A.~Stanciu and C.~Gerigan, ``Comparison between implementations efficiency of
  {HLS} and {HDL} using operations over {Galois} {Fields},'' in \emph{2017
  {IEEE} 23rd {International} {Symposium} for {Design} and {Technology} in
  {Electronic} {Packaging} ({SIITME})}.\hskip 1em plus 0.5em minus 0.4em\relax
  Constanta, Romania: IEEE, Oct. 2017, pp. 171--174. [Online]. Available:
  \url{http://ieeexplore.ieee.org/document/8259883/}
\BIBentrySTDinterwordspacing

\bibitem{stamoulias_hardware_2017}
I.~Stamoulias, C.~Kachris, and D.~Soudris, ``Hardware accelerators for
  financial applications in {HDL} and {High} {Level} {Synthesis},'' in
  \emph{2017 {International} {Conference} on {Embedded} {Computer} {Systems}:
  {Architectures}, {Modeling}, and {Simulation} ({SAMOS})}, Jul. 2017, pp.
  278--285.

\bibitem{Cornu2011}
A.~Cornu, S.~Derrien, and D.~Lavenier, ``Hls tools for fpga: Faster development
  with better performance,'' in \emph{Reconfigurable Computing: Architectures,
  Tools and Applications. ARC 2011. Lecture Notes in Computer Science, vol
  6578. Springer, Berlin, Heidelberg}, 2011.

\bibitem{Bachrach2012}
J.~{Bachrach}, H.~{Vo}, B.~{Richards}, Y.~{Lee}, A.~{Waterman},
  R.~{Avižienis}, J.~{Wawrzynek}, and K.~{Asanović}, ``Chisel: Constructing
  hardware in a scala embedded language,'' in \emph{DAC Design Automation
  Conference 2012}, 2012, pp. 1212--1221.

\bibitem{arcas2014}
O.~{Arcas-Abella}, G.~{Ndu}, N.~{Sonmez}, M.~{Ghasempour}, A.~{Armejach},
  J.~{Navaridas}, {Wei Song}, J.~{Mawer}, A.~{Cristal}, and M.~{Luján}, ``An
  empirical evaluation of high-level synthesis languages and tools for database
  acceleration,'' in \emph{2014 24th International Conference on Field
  Programmable Logic and Applications (FPL)}, 2014, pp. 1--8.

\bibitem{gurel2016}
M.~Gurel, ``A comparative study between rtl and hls for image processing
  applications with fpgas,'' Master's thesis, UNIVERSITY OF CALIFORNIA, SAN
  DIEGO, UC San Diego, 2016, https://escholarship.org/uc/item/9vx1s37b.

\bibitem{pelcat_design_2016}
\BIBentryALTinterwordspacing
M.~Pelcat, C.~Bourrasset, L.~Maggiani, and F.~Berry,
  ``\BIBforeignlanguage{en}{Design productivity of a high level synthesis
  compiler versus {HDL}},'' in \emph{\BIBforeignlanguage{en}{2016
  {International} {Conference} on {Embedded} {Computer} {Systems}:
  {Architectures}, {Modeling} and {Simulation} ({SAMOS})}}.\hskip 1em plus
  0.5em minus 0.4em\relax Agios Konstantinos, Samos Island, Greece: IEEE, Jul.
  2016, pp. 140--147. [Online]. Available:
  \url{http://ieeexplore.ieee.org/document/7818341/}
\BIBentrySTDinterwordspacing

\bibitem{tetrault_two_2018}
M.-A. Tétrault, ``\BIBforeignlanguage{en}{Two {FPGA} {Case} {Studies}
  {Comparing} {High} {Level} {Synthesis} and {Manual} {HDL} for {HEP}
  applications},'' in \emph{\BIBforeignlanguage{en}{2018 21st IEEE NPSS Real
  Time Conference}}.\hskip 1em plus 0.5em minus 0.4em\relax Williamsburg: IEEE,
  Jun. 2018, p.~3.

\bibitem{akkad2018}
G.~{Akkad}, A.~{Mansour}, B.~{ElHassan}, F.~L. {Roy}, and M.~{Najem}, ``Fft
  radix-2 and radix-4 fpga acceleration techniques using hls and hdl for
  digital communication systems,'' in \emph{2018 IEEE International
  Multidisciplinary Conference on Engineering Technology (IMCET)}, 2018, pp.
  1--5.

\bibitem{aledo_vhdl_2019}
\BIBentryALTinterwordspacing
D.~Aledo, B.~Carrion~Schafer, and F.~Moreno, ``\BIBforeignlanguage{en}{{VHDL}
  vs. {SystemC}: {Design} of {Highly} {Parameterizable} {Artificial} {Neural}
  {Networks}},'' \emph{\BIBforeignlanguage{en}{IEICE Transactions on
  Information and Systems}}, vol. E102.D, no.~3, pp. 512--521, Mar. 2019.
  [Online]. Available: \url{http://dx.doi.org/10.1587/transinf.2018EDP7142}
\BIBentrySTDinterwordspacing

\bibitem{millonCASE2020}
R.~Millon, E.~Frati, and E.~Rucci, ``{Implementación de Filtro de Detección
  de Bordes Sobel en SoC usando Síntesis de Alto Nivel},'' in \emph{Actas del
  Congreso Argentino de Sistemas Embebidos (CASE 2020)}, 2020, pp. 73--75.

\bibitem{nosrat_hardware_2012}
\BIBentryALTinterwordspacing
A.~Nosrat and Y.~S.~Kavian, ``Hardware description of multi-directional fast
  sobel edge detection processor by {VHDL} for implementing on {FPGA},''
  \emph{International Journal of Computer Applications}, vol.~47, no.~25, pp.
  1--7, 2012. [Online]. Available:
  \url{http://research.ijcaonline.org/volume47/number25/pxc3879872.pdf}
\BIBentrySTDinterwordspacing

\bibitem{mehra_area_2012}
\BIBentryALTinterwordspacing
R.~Mehra and R.~Verma, ``Area efficient {FPGA} implementation of sobel edge
  detector for image processing applications,'' \emph{International Journal of
  Computer Applications}, vol.~56, no.~16, pp. 7--11, 2012. [Online].
  Available:
  \url{http://research.ijcaonline.org/volume56/number16/pxc3883086.pdf}
\BIBentrySTDinterwordspacing

\bibitem{chaple_design_2014}
G.~Chaple and R.~D. Daruwala, ``Design of {Sobel} operator based image edge
  detection algorithm on {FPGA},'' in \emph{2014 International Conference on
  Communication and Signal Processing}.\hskip 1em plus 0.5em minus 0.4em\relax
  Melmaruvathur, India: IEEE, apr 2014.

\bibitem{bhatt_analysis_2012}
K.~Bhatt, V.~Tarey, and P.~Patel, ``Analysis of source lines of code({SLOC})
  metric,'' \emph{nternational Journal of Emerging Technology and Advanced
  Engineering}, vol.~2, 2012.

\end{thebibliography}

\end{document}